# Photo-excitation band-structure engineering of 2$H$-$NbSe_2$ probed by time- and angle-resolved photoemission spectroscopy

Mari Watanabe[1,2], Takeshi Suzuki[2,*], Takashi Someya[2], Yu Ogawa[2], Shoya Michimae[2], Masami Fujisawa[2], Teruto Kanai[2], Jiro Itatani[2], Tomohiko Saitoh[1], Shik Shin[3,4], and Kozo Okazaki[2,4,5,†]

[1]*Department of Applied Physics, Graduate School of Science, Tokyo University of Science, Katsushika-ku, Tokyo 125-8585 Japan*

[2]*The Institute for Solid State Physics, The University of Tokyo, Kashiwa, Chiba 277-8581 Japan*

[3]*Office of University Professor, The University of Tokyo, Kashiwa, Chiba 277-8581 Japan*

[4]*Material Innovation Research Center, The University of Tokyo, Kashiwa, Chiba 277-8561, Japan*

[5]*Trans-scale Quantum Science Institute, The University of Tokyo, Bunkyo-ku, Tokyo 113-0033, Japan*

We investigated the nonequilibrium electronic structure of 2$H$-$NbSe_2$ by time- and angle-resolved photoemission spectroscopy. We find that the band structure is distinctively modulated by strong photo-excitation, as indicated by the unusual increase in the photoelectron intensities around $E_F$. In order to gain insight into the observed photo-induced electronic state, we performed DFT calculations with modulated lattice structures, and found that the variation of the Se height from the Nb layer results in a significant change in the effective mass and band gap energy. We further study the momentum-dependent carrier dynamics. The results suggest that the relaxation is faster at the K-centered Fermi surface than at the Γ-centered Fermi surface, which can be attributed to the stronger electron-lattice coupling at the K-centered Fermi surface. Our demonstration of band structure engineering suggests a new role for light as a tool for controlling the functionalities of solid-state materials.

Engineering the band structure is a key method for changing the physical properties of solid-state materials. For example, increasing the effective mass significantly increases the electronic heat capacity and magnetic susceptibility [1] [2], while controlling the size of the band gap can tune the absorption and emission spectra [3]. These modifications have conventionally been performed by elemental substitution or applying physical pressure. Photo-excitation has substantial advantages over these methods because the physical properties of materials can be changed without any physical contact or further fabrication [4] [5] [6]. Furthermore, by using ultrashort pulses, material functionalities can be immediately activated. Therefore, photo-excitation has substantial potential for future applications in ultrafast electrical switching [7] [8] [9] [10].

The family of transition metal dichalcogenides (TMDCs) provides rich physics as materials for investigation due to the subtle balance between multiple competing orders such as the Mott insulator, charge density wave (CDW), and superconducting states [11]. This balance can be fairly easily broken by physical pressure [12] [13] or processing into few-layer structures [14] [15]. Photo-excitation has also been frequently employed to reveal hidden properties which do not appear under equilibrium conditions [16].

Among the TMDC materials, 2$H$-$NbSe_2$ has attracted significant interest because it possesses a unique profile in which the CDW and superconducting orders coexist. Figure 1(a) shows the crystal structure of 2$H$-$NbSe_2$. At room temperature, 2$H$-$NbSe_2$ is metallic owing to its half-filled $dz^2$ band. With decreasing temperature, it undergoes a CDW transition at 33 K [17] with the CDW vector, $\boldsymbol{q}$, of 2/3ΓM [18], and a superconducting transition at 7 K [19]. With regard to the CDW phase, two previous static angle-resolved photoemission spectroscopy (ARPSE) studies have reported different momentum dependences of the CDW gap size. One study showed that the CDW gap size is the largest along the MK direction and the smallest along ΓK [20]. This observation is not consistent with the nesting of the Fermi surface (FS), and suggests the possibility of another origin for the CDW gap anisotropy in 2$H$-$NbSe_2$. In contrast, another study reported that the CDW gap opens along ΓK [21]. Thus, the origin of the CDW in 2$H$-$NbSe_2$ remains unclear. These contradicting reports indicate that another approach beyond static ARPES is required.

A non-equilibrium profile obtained using pump-probe type measurements can offer richer information not accessible by static measurements [22] [23]. Time- and angle-resolved photoemission spectroscopy (TARPES) is a very powerful technique for investigating momentum-dependent non-equilibrium behavior which can temporally track the photo-excited

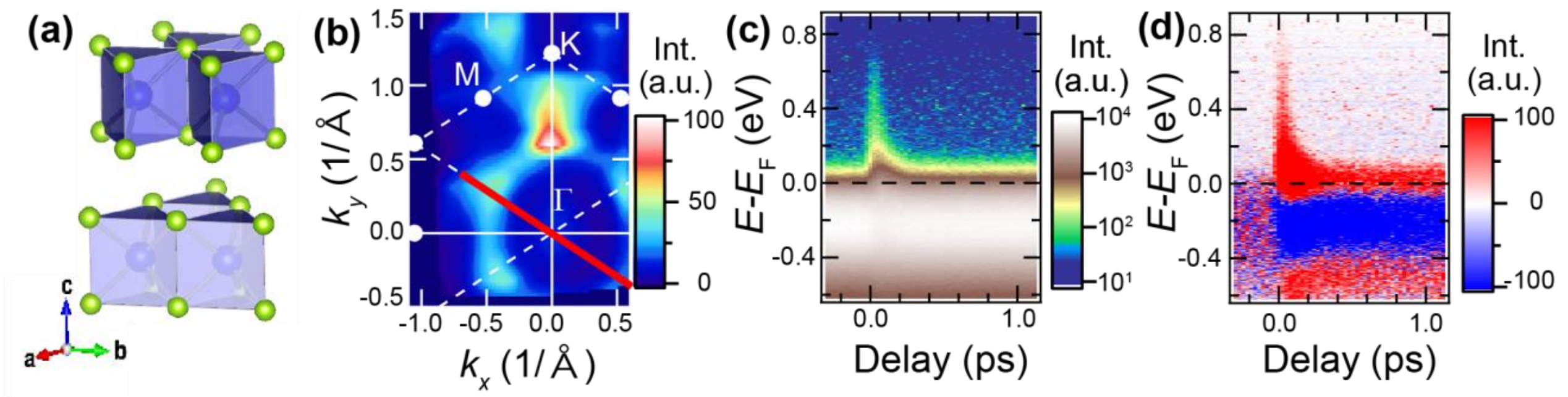

**FIG. 1 (a)** Crystal structure of 2*H*-$NbSe_2$. **(b)** ARPES intensity map measured from a He discharge lamp ($h\nu$ = 21 eV). The intensity is integrated within ±10 meV from $E_F$. **(c)** Row and **(d)** difference spectra of time-dependent energy distribution curves (EDCs). The EDCs are integrated along the red line in Fig. 1**(a)**.

electronic band structure [24] [25] [10]. This technique has been successfully applied to other TMDC systems in which the time-domain observations revealed the excitonic nature of 1*T*-$TiSe_2$ [26] and the insulating nature of 1*T*-$TaS_2$ [27].

In this work, we use TARPES to study the nonequilibrium electronic states in 2*H*-$NbSe_2$ driven by photo-excitations. We find that photo-excitation induces a significant change in the effective mass and band gap energy. To confirm that the observed change in the electronic band structure is caused by the displacive excitation of coherent phonon (DECP) mechanism [28], we perform DFT calculations with modulated lattice structures. Furthermore, we study the electron-lattice couplings by measuring the momentum dependence of the electronic dynamical properties. The results suggest stronger electron-lattice couplings on the K-centered FS.

High-quality single crystals of 2*H*-$NbSe_2$ were purchased from HG Graphene. We confirmed the orientation of the samples by static ARPES. The orientation is consistent with the shape of the samples. For the static ARPES, we used a photon energy of 21.2 eV obtained from a He-I lamp (Scienta Omicron VUV 5050), and detected the photoelectrons using a Scienta R4000 analyzer. In the TARPES setup, we used a Ti:sapphire laser system (Coherent Astrella) with a pulse width of 35 fs, center wavelength of 800 nm (1.55 eV), and repetition rate of 1 kHz. The laser pulses were split into pump and probe pulses. For the pump pulses, we used the fundamental wavelength and set the fluence to 2 mJ/cm$^2$. For the probe pulses, we first doubled the photon energy to 3.10 eV using a $\beta$-$BaB_2O_4$ crystal, and then focused the pulses on the Ar gas filled in the gas cell to achieve high-harmonic generation. We selected the 9th harmonic corresponding to 27.9 eV using a pair of SiC/Mg multilayer mirrors. By using such a high photon energy for the probe pulses, the entire Brillouin zone of 2*H*-$NbSe_2$ can be captured. We estimated the temporal resolution to be less than 80 fs by measuring the response time of highly-oriented pyrolytic graphite and an energy resolution of 250 meV by measuring the energy distribution curves (EDCs) for bulk gold. All the measurements in this study were carried out at the temperature of 15 K unless otherwise stated.

Band structure calculations based on density functional theory (DFT) were performed using the WIEN2k package [29]. $a$ = 3.45 Å and $c$ = 12.54 Å were used for the lattice parameters in the equilibrium state [11].We first performed static ARPES measurements on 2*H*-$NbSe_2$ to confirm the orientation of the cleaved surface. Figure 1(b) shows the results of the FS mapping. The integrated energy window is ±10 meV from the Fermi level ($E_F$). The clear six-fold symmetry of the FS allows us to discern the specific symmetry points in momentum space shown in Fig. 1(b).

After confirming the sample orientation by static ARPES, we investigated the relaxation dynamics of the photo-excited electrons at the Γ-centered FS. Figure 1(c) shows the temporal evolution of the momentum-integrated energy distribution curves (EDCs) integrated along the ΓK direction indicated by the red line in Fig. 1(b). After the arrival of the pump pulses, electrons are immediately excited above $E_F$ and relax to their original state. To provide more clarity, Fig. 1(d) shows the difference spectrum obtained by subtracting the average spectrum before the arrival of pump pulses from the EDCs. The red and blue colors correspond to increases and decreases in the intensities of the photoelectrons, respectively. It should be noted that in addition to the rapid rise and relaxation dynamics far above $E_F$, the intensity around $E_F$ also increases after photo-excitation. Furthermore, the intensity at 0.5 eV *below* $E_F$ also increases. This is quite striking because the electrons are transferred from the occupied to unoccupied states by photo-excitation, and thus the intensity of photoelectrons below $E_F$ should generally decrease after the photo-excitation. This strange behavior indicates a change in the density of states (DOS) as a result of the change in the band structures. Additionally, this change persists for longer than 1 ps. Considering previous reports on the fairly short lifetime of approximately 1 ps of the photo-excited state in a metallic system [30], the observed slow relaxation dynamics show that the photo-induced state in 2*H*-$NbSe_2$ is unusually stable (meta-

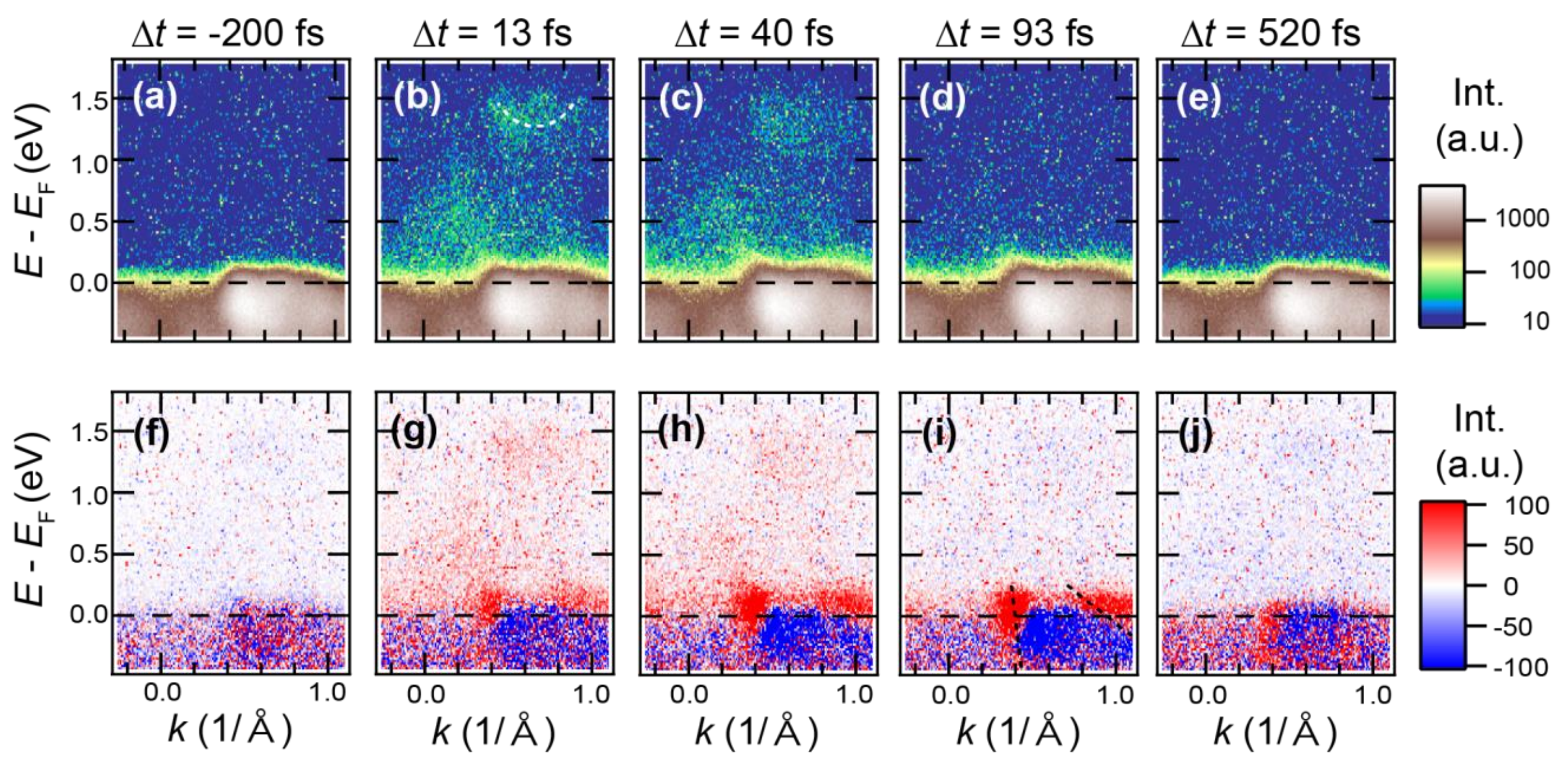

**FIG. 2 (a)–(e)** Row and **(f) –(j)** difference ARPES snapshots at the delay times indicated above. The time delays at which **(f)–(j)** were measured correspond to those in **(a)–(e)** respectively. All data were measured along the ΓK direction.

stable).

To investigate the origin of the unique observed change in the DOS and the long lifetime of the photo-induced state, we measured the momentum-resolved time-dependent band structure along the same cut in energy and momentum space. Figures 2(a)–2(e) show temporal snapshots of the *E*-*k* map at the indicated delay times (Δ*t*). The difference spectra are also shown in Figs. 2(f)–2(j), where red and blue correspond to increases and decreases in the photoelectron intensities, respectively. Before the arrival of the pump (Δ*t* = -200 fs), the hole-like Nb 4*d* and Se 4*p* bands are clearly observed. Just after the pump at Δ*t* = 13.3 and 40 fs, the unoccupied electron bands at around $E$ - $E_F$ = 1.3 eV become populated, as indicated by the white dashed curve in Fig. 2(b). This population is transiently observed and disappears at 93 fs. Focusing on the occupied bands, the original hole-like band is split, as indicated by the two red regions in Figs. 2(g)–2(j). This striking change is most evident at Δ*t* = 93 fs and is indicated by the two black dashed curves in Fig. 2(i).

We now discuss the observed changes in the band structure. Figures 3(a) and 3(b) show the *E*-*k* maps before and after photo-excitation as the difference spectra, respectively. It can be clearly seen that the hole band separates after photo-excitation and appears as a double-sided red region. This observation is most often explained as a consequence of the thermal effects due to the excitation pump. If the electronic temperature in 2*H*-NbSe2 increases above the CDW transition temperature of 33 K, the material would show a phase transition from the CDW phase to the normal metallic phase. In this case, the band structure should show a change similar to that of the phase transition. However, according to previous studies [20] [21], the observed photo-induced state is completely different from the high-temperature band structure, which does not accompany the observed photo-induced change. We further performed TARPES measurements at a higher temperature of 40 K (>$T_{CDW}$), as shown in Fig. S1 [31]. Surprisingly, we observed qualitatively similar behavior to that at 15 K, which suggests that the temperature and CDW phase are largely irrelevant to the observed band changes. Thus, the thermal effect is not a plausible explanation for the observed state.

We suggest an alternative explanation whereby the rearrangement of the electron and lattice structure, as described by the DECP mechanism, is the driving force for the observed photo-induced state. In this rearrangement, near-infrared pulses first transfer the electronic population from the occupied to unoccupied states, and then cause lattice modulation. The electronic band structures are in turn modified and become coordinated with the modulated lattice structure. The absence of coherent phonons in our measurement is explained by the insufficient time resolution of our setup, which is approximately 80 fs. The frequency of coherent phonons corresponding to the stretching mode of Se atoms is 7 THz [32] [33], and thus the time resolution should be better than 35 fs, corresponding to a quarter of the oscillation period. Another reason for the absence of coherent phonons is their strong damping, which prevents their observation over multiple cycles.

To investigate the photo-induced electronic band structure caused by the lattice modulation corresponding to the $A_{1g}$ phonon, we performed DFT calculations within the generalized gradient approximation. Figure 3(c) shows the calculated band structure of 2*H*-$NbSe_2$ with the structure at equilibrium. The Nb 4*d* and Se 4*p* bands lie close to each other along the ΓK direction. To see how this electronic band structure is affected by lattice modulation, we performed DFT calculations on structures with various heights of the Se away from the nearest Nb layer. Figure 3(d) shows the calculation results for the modulated structures schematically shown in Fig. 3(e). It should be noted that the effective mass of the Se 4*p* valence bands changes significantly with increasing Se height. The mechanism behind the observed photo-induced evolution of the TARPES spectra can be understood based on these calculation results. Photo-excitation can change the height of Se in line with the DECP mechanism. This lattice change leads to an increase in the effective mass of the Se *4p* valence band, which results in the separation of the Se *4p* band from the Nb *4d* band, as shown in Figs. 2(i) and 3(b), and an increase in the DOS around $E_F$, as shown in Fig. 1(d). It is also noted that the unoccupied bands shift toward lower energies at the larger lattice modulations shown as circles in Fig. 3(d). This change suggests that the transient population in the unoccupied band shown as the dashed curve in Fig. 2(b) can be considered to be a result of the decrease in the band gap caused by the photo-induced change of the lattice structure.

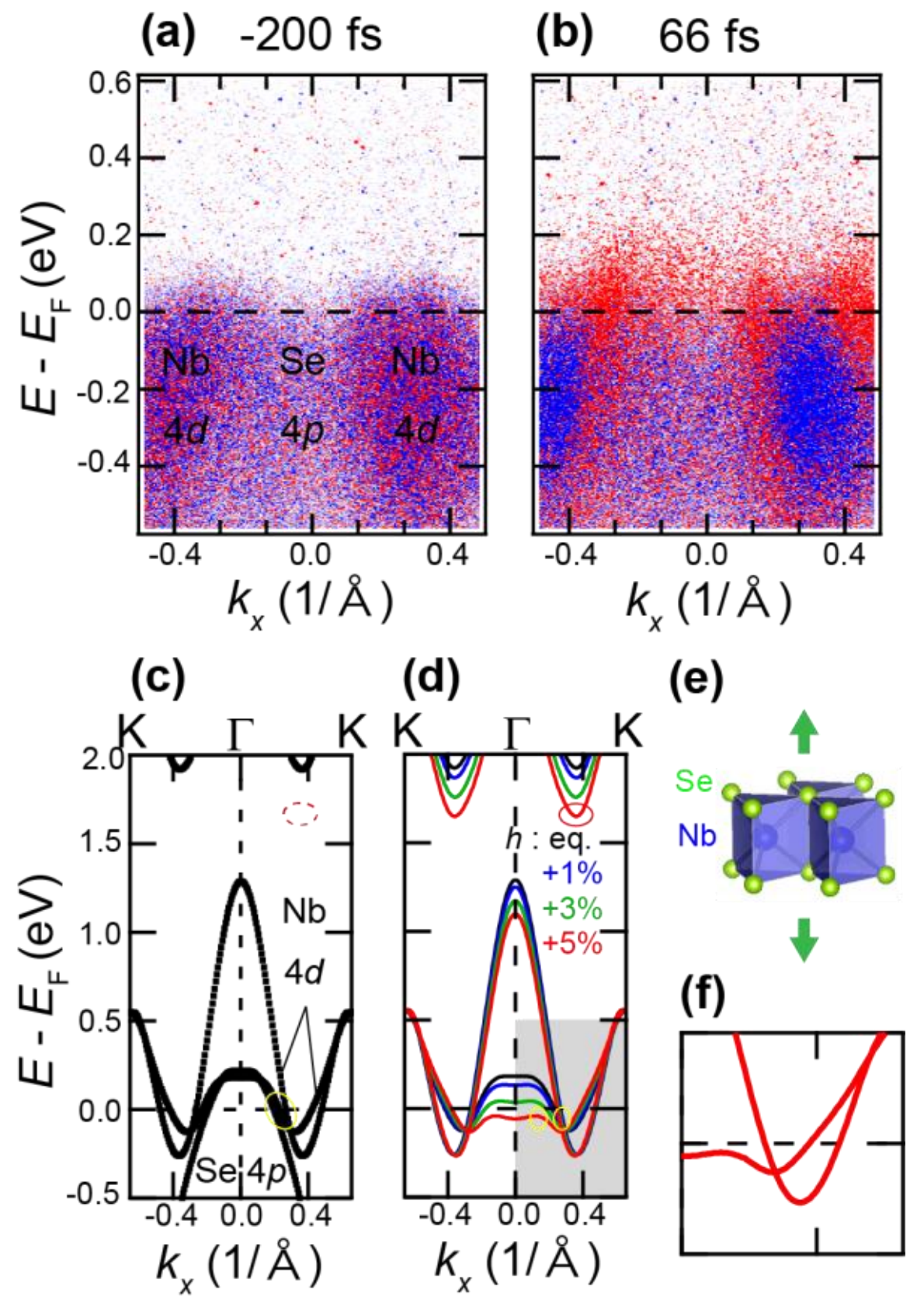

**FIG. 3** ARPES spectra **(a)** before optical pumping and **(b)** at 66 fs. **(c)** Equilibrium band structure via DFT-GGA calculations. **(d)** Variation of band structure due to lattice modulation shown in **(e)**. **(f)** Magnified band structure corresponding to the area indicated by the gray hatch in **(d)**.

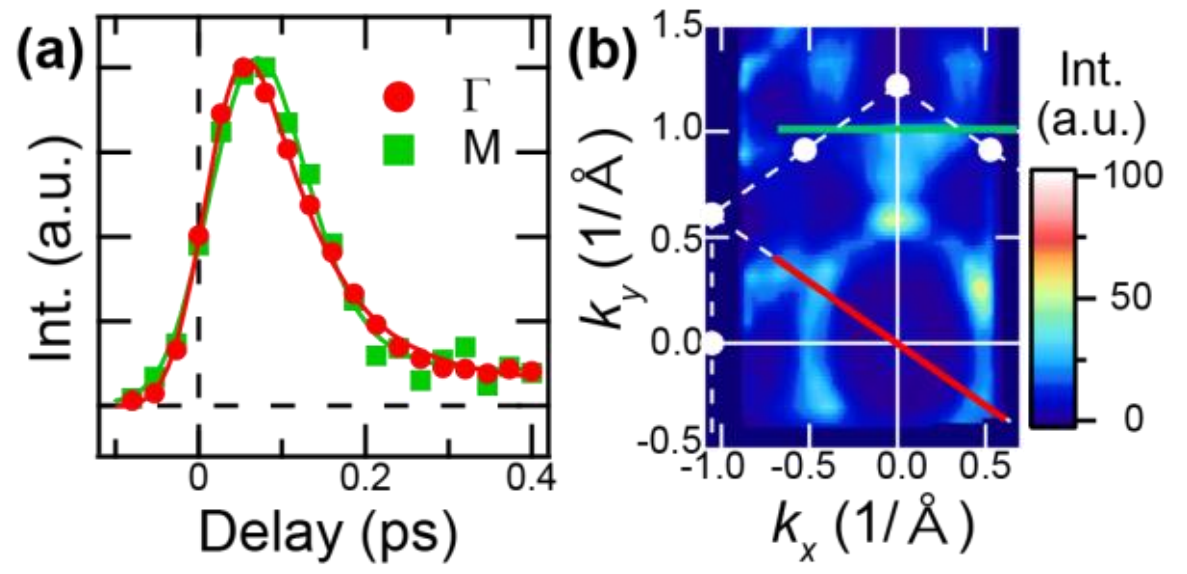

**FIG. 4** **(a)** Time dependence of the intensity integrated along the red and green lines in **(b)**. The markers are the experimental data while the lines are fits. **(b)** Fermi surface map. The energy window is set to be from $E_F$ to + 0.8 eV.

Finally, we discuss the momentum-dependent dynamical behavior and infer its relationship to the CDW gap anisotropy. To investigate the momentum-dependent non-equilibrium properties, we measured the time-dependent photoemission intensities of the FSs at the Γ and K points, and compared the results. Figure 4(a) shows the time-dependent momentum-integrated photoemission intensities of the FSs at the Γ and K points as the red and green markers, respectively. The momentum cuts for the FSs at the Γ and K points are shown as red and green solid lines in Fig. 4(b). To gain a quantitative insight, we fit the experimental data by a single exponential decay function convoluted with a Gaussian function. The fits are shown as red and green solid lines for the FSs at the Γ and K points, respectively. The decay constants of the FSs at the Γ and K points are 78 and 46 fs, respectively. According to the two-temperature model, the pump pulse injects excess energy into the electron system. The energy is then transferred to the lattice system via electron-lattice coupling. The speed of the energy transfer is determined by the electron-lattice coupling constant. The larger the electron-lattice coupling constant, the faster the excess energy is transferred to the lattice system. Our finding of the faster decay of the photoemission intensities of the FS at the K point, which reflects the faster hot carrier dynamics, indicates a stronger electron-lattice coupling.

With regard to the CDW gap anisotropy, the stronger electron-lattice coupling at the K point favors the result of Ref. [21]. However, the carrier dynamics are affected by many factors other than the amplitude of the gap, such as the size of the FS and the populations of the involved phonons. Further studies following the present work will be required.

We studied the photo-induced electronic states in 2*H*-$NbSe_2$ by TARPES. We found distinctive changes in the electronic profile manifested as the persistent increase of photoelectrons around $E_F$ and the separation of the Se *4p* band from the Nb *4d* band as a result of an increase in the effective mass of the Se *4p* band. This unusual state is induced by the lattice modulation in line with the DECP mechanism, and confirmed by DFT calculations. We further investigated the momentum-dependent dynamics. The dynamics indicate a larger electron-lattice coupling at the K point.

We would like to thank Editage (www.editage.com) for English language editing. This work was supported by Grants-in-Aid for Scientific Research (KAKENHI) (Grant No. JP18K13498, JP19H01818, JP19H00651) from the Japan Society for the Promotion of Science (JPSJ), by JSPS KAKENHI on Innovative Areas "Quantum Liquid Crystals" (Grant No. JP19H05826), by the Center of Innovation Program from the Japan Science and Technology Agency, JST, and by MEXT Quantum Leap Flagship Program (MEXT Q-LEAP) (Grant No. JPMXS0118068681), Japan.

*Corresponding author.
takeshi.suzuki@issp.u-tokyo.ac.jp
†Corresponding author.
okazaki@issp.u-tokyo.ac.jp

# Supplementary Material

## Time- and angle-resolved photoemission spectra at 40 K

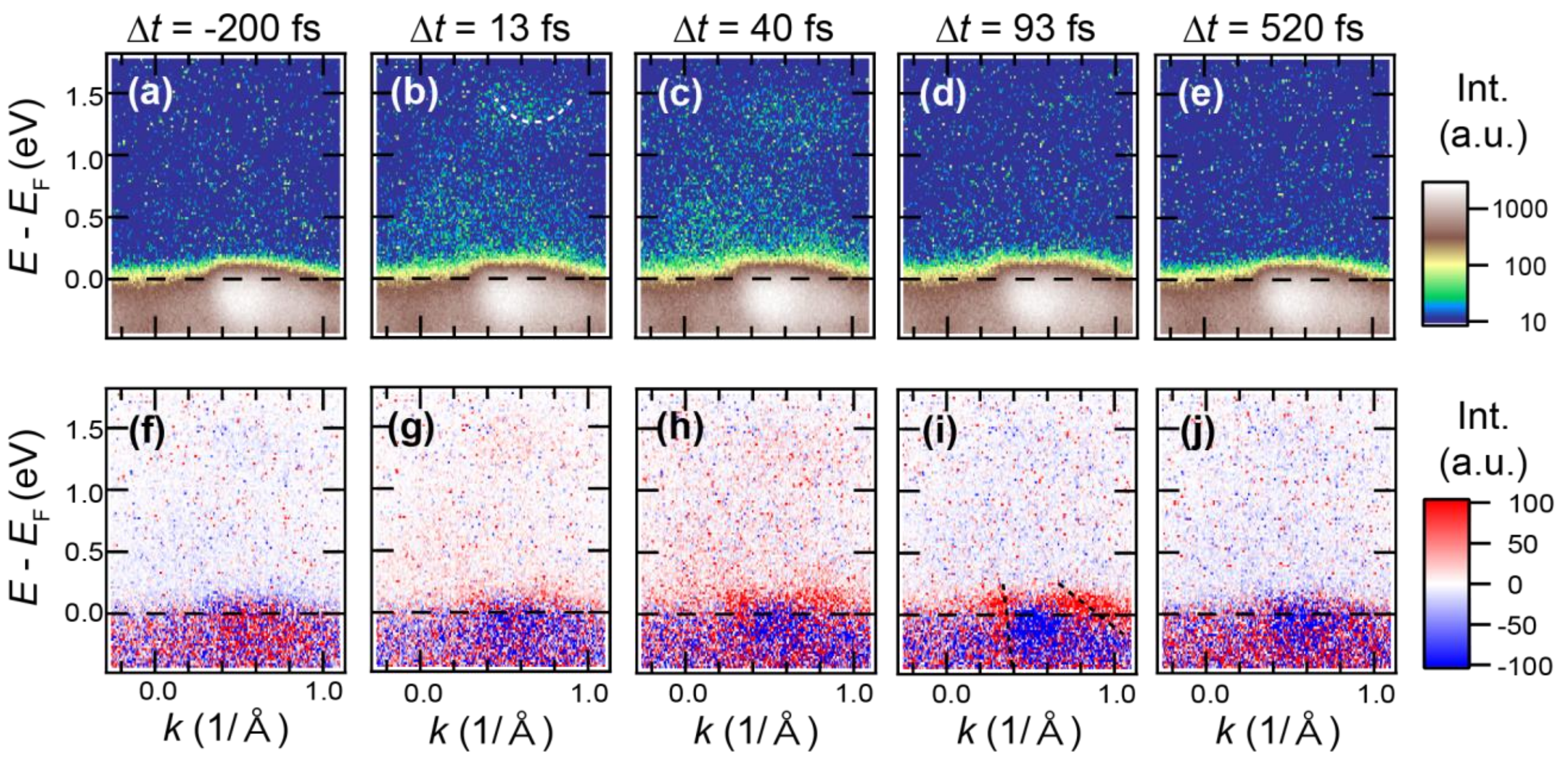

**FIG. S1 (a)-(e)** Row and **(f)-(j)** difference ARPES snapshots at the delay times indicated above. Time delays at which **(f)-(j)** have been measured correspond to **(a)-(e)** respectively. All data have been measured along the ΓK direction and at the temperature of 40 K.